\documentclass[twocolumn,showpacs,preprintnumbers,amsmath,amssymb]{revtex4}

\usepackage{graphicx}
\usepackage{dcolumn}
\usepackage{bm}
\usepackage{quantum}

\def\psihat{\hat{\psi}^\dagger}
\def\zerodag{\hat{0}^\dagger}
\def\onedag{\hat{1}^\dagger}
\def\xdag{\hat{x}^\dagger}

\begin{document}

\preprint{APS/123-QED}

\title{Security of differential phase shift quantum key distribution against individual attacks}

\author{Edo Waks}
\affiliation{
E.L. Ginzton Labs\\
Stanford University, Stanford, CA, 94305
}%

\author{Hiroki Takesue}
\affiliation{
NTT Basic Research Laboratories, NTT corporation\\
3-1 Morinosato Wakamiya, Atsugi, Kanagawa, Japan
}%

\author{Yoshihisa Yamamoto}
\affiliation{
E.L. Ginzton Labs\\
Stanford University, Stanford, CA, 94305
}%
\affiliation{ National Institute of Informatics, Tokyo, Japan}%

\date{\today}

\begin{abstract}
We derive a proof of security for the Differential Phase Shift
Quantum Key Distribution (DPSQKD) protocol under the assumption
that Eve is restricted to individual attacks. The security proof
is derived by bounding the average collision probability, which
leads directly to a bound on Eve's mutual information on the final
key. The security proof applies to realistic sources based on
pulsed coherent light.  We then compare individual attacks to
sequential attacks and show that individual attacks are more
powerful.
\end{abstract}

\pacs{Valid PACS appear here}
\maketitle

\section{\label{sec:Introduction} Introduction}

The goal of quantum cryptography is to exchange an unconditionally
secure secret key over a potentially hostile environment.  To
date, a variety of protocols have been proposed to accomplish this
goal.  The first of these protocols was originally proposed by
Bennett and Brassard (BB84)~\cite{BennettBrassard84}.  Since that
ground-breaking result, a variety of additional protocols have
been
proposed~\cite{Bennett92,Ekert91,BennettBrassard92,KoashiImoto97,HuttnerImoto95},
with varying advantages and disadvantages.

One of the more recent protocols is known as Differential Phase
Shift Quantum Key Distribution (DPSQKD for
short)~\cite{InoueWaks02}. This protocol appears to have several
important advantages which make it extremely promising for
practical systems.  First, DPSQKD can be easily implemented in
optical fibers using readily available optical telecommunication
tools.  Second, there is good indication that DPSQKD is largely
insensitive to multiphoton states generated by the source, as
opposed to other protocols such as BB84.  This allows the
communicating parties to transmit much brighter coherent states,
leading to higher communication rates and longer communication
distances.

To date, all security statements about DPSQKD have been based on
considering only very restricted types of eavesdropping attacks,
such as intercept and resend or inserting a beamsplitter. This
leads to the possibility that more sophisticated attacks based on
generalized quantum measurements may exist which could potentially
nullify many of the advantages of DPSQKD.  Thus, it is important
to have a security proof for this protocol which works for a more
general class of attacks.  Furthermore, because robustness to
photon splitting attacks is one of the main features of this
protocol, it is important that the proof of security includes
these types of attacks.

The most general attacks that one may consider in quantum
cryptography are known as coherent or joint attacks.  In these
types of attacks Eve treats the entire key as a single quantum
system, which is entangled with a probe state.  The probe is only
measured after all classical information is exchanged. Coherent
attacks allow Eve to take advantage of correlations induced by
classical information exchanged during error correction and
privacy amplification.  The proof of security against coherent
attacks is extremely difficult.  To date, there are several proofs
of security for the BB84 protocol against these most general types
of attacks~\cite{Mayers01,ShorPreskill00}.  A general security
proof for the B92~\cite{Bennett92} protocol has also been
derived~\cite{Koashi04}. In order to make the problem more
tractable, one often restricts eavesdropping to individual
attacks.  In these types of attacks, it is assumed that Eve
attaches an independent probe to each photon, and then measures
the probes independently. The security of BB84 against individual
attacks has been investigated in several
works~\cite{Lutkenhaus99,Fuchs97,Slutsky98}.  The security of the
B92 protocol against individual attacks has also been
proven~\cite{TamakiKoashi03}.  The restriction to individual
attacks is often considered a realistic assumption because the
capability to perform joint attacks is well beyond the domain of
modern technology.  Such attacks would require that an
eavesdropper possess a probe of extremely large dimensionality (on
the order of the length of the string) with indefinite coherence
time, and process the probe states with a quantum computer.  Even
individual attacks require a degree of quantum computational power
which seems out of reach for the foreseeable future.

In this paper, we derive a proof of security for DPSQKD against
individual attacks.  The proof applies to realistic sources based
on attenuated lasers, and accounts for the poisson nature of the
photon statistics injected into the channel. Security is proved by
deriving a bound on Eve's average collision probability, which
directly leads to a bound on her mutual information for the final
key~\cite{BennettBrassard95}. We use this result to calculate the
communication rate of DPSQKD in the limit of large strings. We
then compare this rate to that of BB84 using both single photon
sources and poisson light sources. We show that DPSQKD achieves
rates very close to BB84 with an ideal single photon source, and
significantly outperforms BB84 with poisson light.  This is an
important result because DPSQKD requires only attenuated laser
light and linear optics, in contrast to single photon sources
which are difficult to implement.  In the final section of this
paper, we consider another type of eavesdropping attack known as a
sequential attack.  These types of attacks are not individual
attacks, so they are not accounted for by our proof of security.
However, they are conceptually simple and have raised a level of
concern regarding the security of DPSQKD.  We calculate the
communication rate against these types of attacks and compare it
to the rate for individual attacks.  It turns out that in our
parameter range of interest, the communication rate for individual
attacks is always lower than sequential attacks.  Thus security
against individual attacks automatically implies security against
sequential attacks.

\section{Differential Phase Shift QKD}

\begin{figure}
\centering\includegraphics[width=7cm]{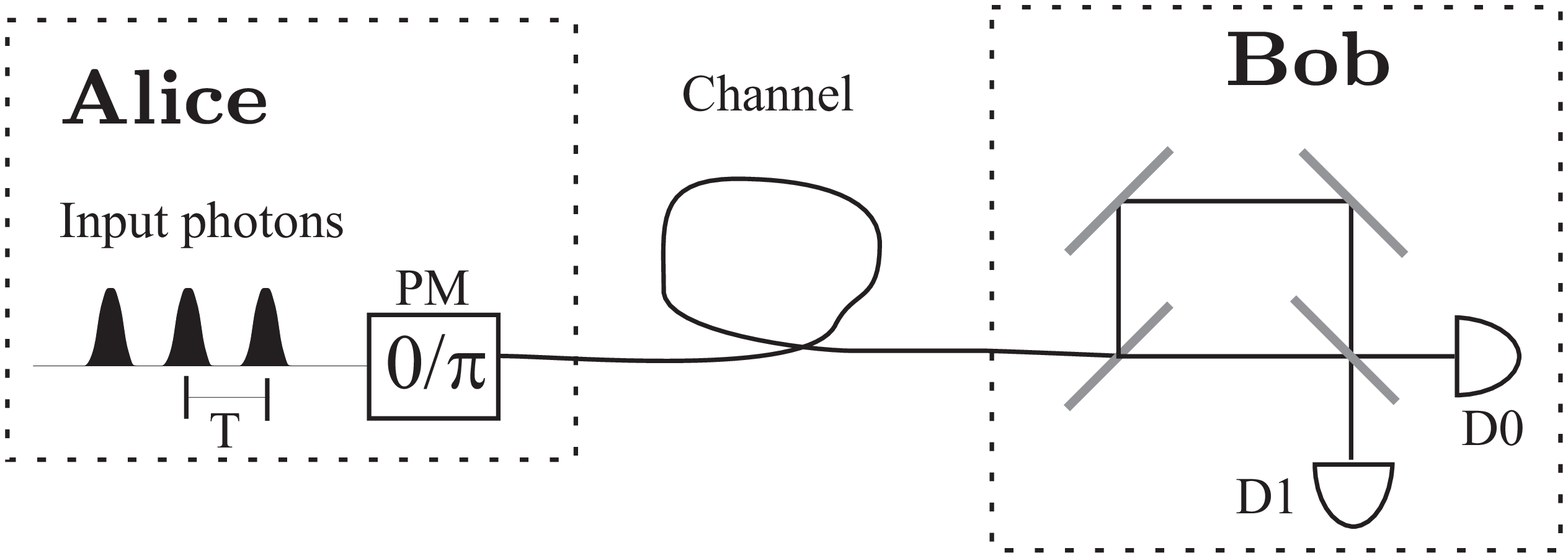} \caption{A
basic DPSQKD system.} \label{fig:DPSQKDfig}
\end{figure}

Figure~\ref{fig:DPSQKDfig} shows the basic idea behind DPSQKD.
Alice prepares a periodic train of attenuated laser pulses whose
phase is randomly modulated to be $0$ or $\pi$.  The coherent
pulses are sent down the quantum channel and received by Bob, who
measures them using an unbalanced interferometer which combines
the partial wave at time slot $n$ with time slot $n+1$ on a
beamsplitter. If the phase difference between these two pulses is
$0$, a detection event will only occur in detector $D0$.
Similarly, if the phase difference is $\pm \pi$, detection events
will only occur in detector $D1$. Bob records the detection events
and the times they occurred at. Once the quantum communication is
done, Bob announces at which times he detected a photon.  This
information allows Eve to determine Bob's string based on her
knowledge of the phase differences.  Error correction and privacy
amplification can then be performed on the sifted key to create
the final secure key.

\begin{figure}
\centering\includegraphics[width=7cm]{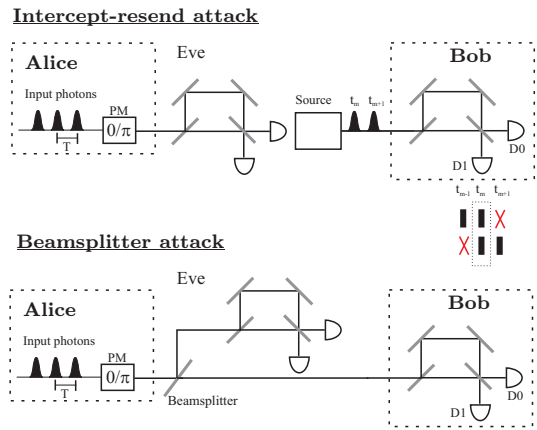}
\caption{Schematic of intercept-resend and beamsplitter
eavesdropping strategies.} \label{fig:IRfig}
\end{figure}

To get an idea as to why this protocol is secure, lets consider
some simple attacks Eve might try to perform.  Two basic attacks
are shown in Figure~\ref{fig:IRfig}.  The first attack is an
intercept and resend strategy, in which Eve uses the same type of
interferometer as Bob.  When Eve gets a detection event time
$t_m$, she learns the phase difference between the pulses at time
$t_m$ and $t_{m+1}$.  She then prepares a pair of pulses with the
measured phase difference and sends them to Bob.  If Bob detects a
photon at time $t_m$, then Eve has successfully stolen a bit
without inducing errors. However, if a detection instead occurs at
times $t_{m+1}$ or $t_{m-1}$, then Bob will observe a $50\%$ error
rate, and Eve will have no knowledge about that bit of the key.
This strategy therefore induces a $25\%$ overall error rate which
can be detected by Alice and Bob, revealing Eve's presence.

In the second strategy, Eve inserts a beamsplitter into the
channel to pull of a fraction of the light.  This split off
fraction is then measured by an unbalanced interferometer, while
the remainder is sent to Bob.  We assume Eve posses a lossless
channel with which she can transmit the un-split photons to Bob.
This allows her to split off a fraction of the photons equal to
the channel loss without modifying the communication rate.
Because coherent states are being used, Eve's detection events are
independent of Bob's.  Thus, the probability that Eve knows the
value of a bit at time $m$ given Bob detected a photon at that
time, denoted $p_e(m)$, is simply given by
  \begin{equation}
    p_{split}(m) = \nbar  \left( 1 - \eta \right) \approx \nbar
  \end{equation}
where $\nbar$ is the average number of photons per pulse.  For
small values of $\nbar$, this attack provides little information
about the sifted key.  If Eve delays her measurement and uses an
optical switch, she can improve the attack a factor of 2.

\section{Photon Splitting in DPSQKD}

In this section we lay the groundwork for the proof of security.
We start by giving a mathematical description of individual
attacks.  We then investigate photon splitting attacks in DPSQKD.
The state prepared by Alice, denote $\|\psi>$, is a set of
consecutive coherent state pulses.  The phase shift $\phi_n$ is
the phase induced by the phase modulator on pulse $n$. This phase
can take on the values $0$ and $\pi$.  If Alice transmits $N$
coherent pulses, we have
  \begin{equation} \label{eq:InitState}
    \|\psi> = \bigotimes_{n=0}^{N-1} \|\alpha e^{i(\phi+\phi_n)}>
  \end{equation}
where $\phi$ is the initial phase of the coherent state. We define
the bosonic operator $\psihat$ as
  \begin{equation}
    \psihat = \frac{1}{\sqrt{N}}\sum_{n=0}^{N-1} e^{i\phi_n}\adag_n
  \end{equation}
where $\adag_n$ is the creation operator for a photon in time slot
$n$.  Assuming that the time slots do not overlap, these different
operators commute with each other.  Thus, the state in
Eq.~\ref{eq:InitState} can be re-written as
  \begin{equation}
    \|\psi> = \sum_{j=0}^{\infty} \sqrt{P(j)}e^{i j \phi} \frac{\left(\psihat\right)^j}{\sqrt{j!}} \|0>
  \end{equation}
where $P(j)$ is a poisson distribution with average photon number
$N\nbar$, and $\nbar=|\alpha|^2$.  A fundamental assumption of the
DPSQKD protocol is that Eve does not possess a phase reference.
Because of this, the above state should be averaged out over the
different values of the phase $\phi$, resulting in the mixed state
  \begin{equation}
    \rho_e = \sum_{j=0}^{\infty} P(j) \kb|\psi_j><\psi_j|
  \end{equation}
where $\|\psi_j> = (\psihat)^j/\sqrt{j!} \|0>$.  With no loss of
generality, Eve can measure the photon number using a state
preserving quantum non-demolition (QND) measurement.  She can then
split off $N\nbar T$ of the photons, where $T$ is the transmission
efficiency of the channel, and send them to Bob, while storing
$N\nbar (1-T)$ photons coherently to be measured after Alice and
Bob have revealed all classical information.

There are now two components of the eavesdropping strategy which
must be addressed.  The first is how much information can be
extracted from the split photons.  This component is analogous to
the information obtained from photon splitting attacks in BB84.
Second, in the presence of channel noise Eve can potentially
attack the fraction of the key that she transmits to Bob by
entangling it with a probe state.  This part of the eavesdropping
attack is analogous to the general POVM attacks on single photon
states.  We will investigate the split photon component first, and
then the generalizes POVM on the transmitted photons.

Our analysis makes an auxiliary assumption that Eve attacks each
photon individually.  For the photons that are transmitted to Bob,
each one is individually split and attached to an independent
probe.  The probes are then independently measured after all
classical communication is received.  The split photons are also
individually stored and measured.  The individual attacks
assumption implies that Eve cannot use the measurement results of
one photon to refine her measurement on the rest of the photons.
Thus, if Eve has split off $k$ photons, she has $k$ copies of the
state $\psihat\|0>$. Eve stores these $k$ copies coherently until
all public information is revealed.  After the quantum
transmission is done, Bob will publicly announce the time slots in
which he had a detection event.  Let $B$ be the set of all time
slots in which a detection event was observed, and $\bar{B}$ be
the set of all other time slots. The operator $\psihat$ can be
re-written as
  \begin{equation} \label{eq:SplitState}
    \psihat = \frac{1}{\sqrt{N}}\left[ \sum_{m\in B}
    e^{i\phi_m}\left(\adag_m+ e^{i\Delta\phi_m} \adag_{m+1}
    \right) + \sum_{n\in \bar{B}}e^{i\phi_n}\adag_n\right]\|0>
  \end{equation}
For each time slot in $B$, Eve can perform the following unitary
transformation
  \begin{eqnarray}
    \adag_m & \to & \frac{1}{\sqrt{2}} \left( \zerodag_m +
    \onedag_m
    \right) \\
    \adag_{m+1} & \to & \frac{1}{\sqrt{2}} \left( \zerodag_m -
    \onedag_m
    \right)
  \end{eqnarray}
where $\zerodag_m$ and $\onedag_m$ are orthogonal modes. There is
no loss of generality in assuming this transformation is
performed, because it is unitary and simply represents a
transformation of the measurement basis. If measurement basis
$\|E>$ is optimal for the state in Eq.~\ref{eq:SplitState}, then
the basis $U^\dagger\|E>$ is now optimal after the unitary
transformation $U$ is applied. The state of each split photon is
now given by
  \begin{equation}
    \psihat = \frac{1}{\sqrt{N}}\left[ \sum_{m\in B}
    e^{i\phi_m}\sqrt{2}\xdag_i + \sum_{n\in \bar{B}}e^{i\phi_n}\adag_n\right]
  \end{equation}
where $\xdag_i$ is $\zerodag_i$ if Alice sent a binary $0$, and
$\onedag_i$ if Alice sent $1$.  Thus, Eve's split photons are in a
linear superposition of all the bits of the secret key, plus the
irrelevant time slots where no photon was detected. However,
because Eve does not know the phases $\phi_m$, her state is in
fact a mixture of the different values of $\phi_m$.  Specifically,
  \begin{eqnarray}
    \rho_e & = & \sum_{\phi_1\ldots\phi_k}
    p(\phi_1,\ldots,\phi_k)\psihat\kb|0><0|\hat{\psi} \nonumber \\
    & = & \frac{1}{N}\left[ 2\sum_{m\in B} \kb|x_m><x_m| +
    \sum_{n\in \bar{B}} \kb|n><n| \right]\label{eq:SplitStateFinal}
  \end{eqnarray}
In the above equation $\|x_m>=\xdag_m\|0>$ and $\|n>=\adag_n\|0>$.
The phases $\phi_i$ are summed over the possible values of $0$ and
$\pi$, which have equal probability so that
$p(\phi_1,\ldots,\phi_k)=1/2^k$.  From
Eq.~\ref{eq:SplitStateFinal} we see that Eve's state is in fact a
random mixture of orthogonal states.  This turns the problem into
one of classical probability theory instead of quantum
measurement. That is, if Bob recorded $y$ detection events, each
split photon will reveal a bit of Eve's key with probability
$2y/N$, and will reveal no information at all with probability
$1-2y/N$.

Let us define $T$ as the channel transmission and $\nbar$ as the
average number of photons per pulse.  After $N$ pulses, Bob will
observe on average $N\nbar T$ detection events.  Assuming Eve has
possession of a lossless channel, she must transmit $N\nbar T$
photons to Bob, and can split off the remainder $N\nbar (1-T)$
photons to be stored coherently.  After Bob reveals the time slots
of his detection events, Eve can measure her split photons, in
which case she learns $2N\nbar^2T(1-T)$.  Thus, from the split
photons Eve learns a fraction $2\nbar(1-T)\approx 2\nbar$ of the
sifted key.  If $\nbar=0.1$, Eve learns only $20\%$ of the final
key.

The most important aspect of the above conclusion is that, in
contrast to BB84, the amount of information Eve obtains from
photon splitting attacks is independent of channel loss.  In BB84,
as the channel losses get larger Eve can preferentially transmit
multi-photon states and block off an appropriate fraction of the
single photon states to conserve the overall communication rate.
As the channel loss becomes larger, this type of attack gives her
complete information over an increasingly larger fraction of the
key.  This results in a final communication rate which is roughly
a quadratic function of channel loss, and hence decreases very
quickly.  In contrast, in DPSQKD the fraction of the final key
that is revealed is only a function of $\nbar$.  This leads to a
communication rate which decreases only linearly with channel
loss, indicating robustness against photon splitting attacks.

\section{Proof of security}

In the previous section we showed that due to photon splitting,
Eve obtains complete information over a fraction $2\nbar$ of the
key.  When $\nbar$ is small, photon splitting attacks are largely
ineffective.  However, in the presence of channel noise Eve can
also attack the photons that she transmits to Bob by entangling
them with a probe state, and then measuring the probe after all
classical information has been revealed.

Because we restrict our attention to individual attacks, it is
assumed that Eve attaches an independent probe to each photon, and
these probes are all measured independently.  The goal of a proof
of security is to come up with a bound for the average collision
probability~\cite{Lutkenhaus99}, defined as
  \begin{equation}
    P_c = \sum_{x,z,m} p^2(X=x|Z=z,M=m)p(z,m)
  \end{equation}
where $X$ is the key Alice transmitted to Bob, $Z$ is the
information Eve obtained from measuring the photon, and $M$ is the
set of time slots in which Bob detected a photon, which is also
known to Eve. For the case of individual attacks, bit $i$
originated from one photon which is correlated to an independent
probe state $Z_i$, as well as $M_i$ which is the time of the
detection. In this case, the collision probability simplifies to a
product of the collision probabilities of each individual
bit~\cite{Lutkenhaus96}. Thus,
  \begin{equation}
    P_c = \prod_i Pc_0
  \end{equation}
where
  \begin{equation} \label{eq:IndPc}
    Pc_0 =\sum_{x,z,m} p^2(X_i=x|Z_i=z,M_i=m)p(Z_i=z,M_i=m)
  \end{equation}
If bit $i$ occurred in a time slot where Eve has obtained its
value due to photon splitting, then $Pc_i=1$.  Let $\bar{S}$ be
the set of all bits that occurred in time slots which do not
coincide with a photon splitting measurement.  We now have
  \begin{equation} \label{eq:TotalPc}
    P_c = \prod_{i\in \bar{S}} Pc_0
  \end{equation}
We adopt a simplified notation such that
$P(X_i=x|Z_i=z,M_i=m)=p(x|z,m)$, and use similar notation for all
other probability distributions.  In appendix~\ref{ap:Pc} we show
that the expression in Eq.~\ref{eq:IndPc} can be re-written as
  \begin{equation} \label{eq:IndPcRearranged}
    Pc_0 = \sum_{m} p(m) \left( 1 - \frac{1}{2p(m)}\sum_{z}
    \frac{p(z,m |0)p(z,m|1)}{p(z,m)} \right)
  \end{equation}
where $0$ and $1$ are the possible values of the bit Alice
transmitted.

We now develop a mathematical formalism for all possible
measurements Eve can perform.  We define $\|E_i>$ as the initial
state of Eve's hilbert space.  We do not assume anything about the
dimensionality of this space.  The initial state of a photon-probe
system is given by
  \begin{equation}  \label{eq:InitHilbert}
    \|\Psi> = \frac{1}{\sqrt{N}} \sum_n e^{i\phi_n} \|n>\|E_i>
  \end{equation}
where $\|n>$ is one again defined as $\adag_n\|0>$ and represents
a photon in time slot $n$.  The most general unitary
transformation Eve can apply to the system is described by
  \begin{equation}
    \|n>\|E_i> \to \sum_m \|m> \|E_{n,m}>
  \end{equation}
where $\|E_{n,m}>$ are states in Eve's Hilbert space and are not
assumed to be normalized or orthogonal.  Plugging the above
relation back into Eq.~\ref{eq:InitHilbert} and rearranging the
summation we obtain
  \begin{eqnarray}
    \|\Psi> & = & \frac{1}{\sqrt{N}} \sum_m \|m> \sum_n
    e^{i\phi_n}\|E_{n,m}> \nonumber \\
    & = & \frac{1}{\sqrt{N}} \sum_m \|m> \|J_m>
  \end{eqnarray}
After Bob's interferometer, the state is once again transformed
into
  \begin{equation}
    \|\Psi> = \frac{1}{2\sqrt{N}} \sum_m \left[ \left(\|J_m> +
    \|J_{m+1}> \right) \|0_m> +\left( \|J_m> - \|J_{m+1}> \right)
    \|1_m> \right]
  \end{equation}
where $\|0_m>$ and $\|1_m>$ represent a photon in the output ports
of Bob's interferometer which correspond to a binary $0$ or $1$ at
time $m$.

In appendix~\ref{ap:ErrRate}, it is shown that the probability of
an error given Bob detected a photon at time $m$ is given by the
expression
  \begin{eqnarray}
    p_{e|m} & = & \frac{1}{2}\left[1 - \frac{1}{Np(m)} \left(
    \bk<E_{m,m}|E_{m+1,m+1}>\right. \right.\nonumber \\
    & & \left. \left.+ \bk<E_{m,m+1}|E_{m+1,m}> \right)\label{eq:ErrRateTot}
    \right]
  \end{eqnarray}
Eve will measure her probe in the basis $\|z>$, which cannot
depend on $\phi_m$ since this information is unavailable.  We
define the number $E_{n,m}(z)= \bk<z|E_{n,m}>$.  Without loss of
generality we can assume this to be a real number.  We do not need
to introduce complex numbers in this case because a probe state
with a complex probability amplitude can always be replaced by a
probe of higher dimensionality with real probability amplitudes
which performs at least as well~\cite{Lutkenhaus99}. We also
define the following expressions:
  \begin{eqnarray}
    Q_m(z) & = & E_{m,m}(z) + E_{m+1,m}(z) \\
    P_m(z) & = & E_{m,m}(z) - E_{m+1,m}(z) \\
    Q_{m+1}(z) & = & E_{m,m+1}(z) + E_{m+1,m+1}(z) \\
    P_{m+1}(z) & = & E_{m,m+1}(z) - E_{m+1,m+1}(z)
  \end{eqnarray}
In appendix~\ref{ap:CollProb} we show that the collision
probability is given by the expression
  \begin{widetext}
  \begin{equation}
    Pc_0 = 1 - \frac{1}{4N}\sum_{m,z} \frac{ \left( Q_m^2(z) +
    Q_{m+1}^2(z) + \sum_{n\neq m,m+1} E_{n,m}^2 +E_{n,m+1}^2
    \right)\left( P_m^2(z) +
    P_{m+1}^2(z) + \sum_{n\neq m,m+1} E_{n,m}^2 +E_{n,m+1}^2
    \right)}{\sum_n E_{n,m}^2 +E_{n,m+1}^2} \label{eq:PcRaw}
  \end{equation}
  \end{widetext}
From the above expressions, it is clear that $E_{n,m}(z)$ where
$n\neq m-1,m,m+1$ can only decrease Eve's collision probability
while simultaneously increasing the error rate.  Thus, we only
need to consider the states $\|E_{m-1,m}>$,$\|E_{m,m}>$, and
$\|E_{m+1,m}>$.  We relabel these states as $\|A_m>$, $\|B_m>$,
and $\|C_m>$ respectively.  We similarly define
$A_m(z)=\bk<z|A_m>$, $B_m(z)=\bk<z|B_m>$, $C_m(z)=\bk<z|C_m>$. The
probability of error is now given by
  \begin{equation} \label{eq:ErrRate}
    p_{e|m} = \frac{1}{2} - \frac{1}{2Np(m)} \sum_m \left( \bk<B_m|B_{m+1}> +
    \bk<C_m|A_{m+1}> \right)
  \end{equation}
We also have the expression
  \begin{eqnarray}
    Q_m(z) & = & B_m(z) + C_m(z) \\
    P_m(z) & = & B_m(z) - C_m(z) \\
    Q_{m+1}(z) & = & A_{m+1}(z) + B_{m+1}(z) \\
    P_{m+1}(z) & = & A_{m+1}(z) - B_{m+1}(z)
  \end{eqnarray}

In appendix~\ref{ap:ColProbBound} it is shown that the collision
probability is upper bounded by
  \begin{eqnarray}
    Pc_0 & \le & 1 - \frac{1}{8N}\sum_{m,z}
    \left( \bk<A_m|A_m> + \bk<C_{m+1}|C_{m+1}>\right. \nonumber \\
    & & +\bk<Q_m|P_m>+\bk<Q_{m+1}|P_{m+1}> \nonumber \\
      & &\left. + \bk<Q_m|P_{m+1}>+\bk<Q_{m+1}|P_{m}>\right)
      \label{eq:PcBound0}
  \end{eqnarray}
In appendix~\ref{ap:SymAttack} we show that there is always an
optimal attacks which satisfies the property that the inner
product of the vectors $\|A_m>$, $\|B_m>$, and $\|C_m>$ with any
other vector from this set is independent of $m$.  This directly
implies that $p(m)=1/N$ and that the collision probability is
independent of $m$.  Thus,
  \begin{eqnarray}
    Pc_0 & \le & 1 - \frac{1}{8}\sum_{z}
    \left( \bk<A_0|A_0> + \bk<C_{1}|C_{1}> \right. \nonumber \\
    & & +\bk<Q_0|P_0>+\bk<Q_{1}|P_{1}>  \nonumber \\
    & & \left. +
    \bk<Q_0|P_{1}>+\bk<Q_{1}|P_{0}>\right) \label{eq:PcBound}\\
    e & = & \frac{1-\bk<B_0|B_{1}> - \label{eq:constraint}
    \bk<C_0|A_{1}>}{2}
  \end{eqnarray}
where $e$ is the bit error rate of the transmission. We must now
maximize Eq.~\ref{eq:PcBound} subject to the constraint in
Eq.~\ref{eq:constraint}.  This is done in
appendix~\ref{ap:optimization}, where it is shown that
  \begin{equation} \label{eq:FinalPcBound}
    Pc_0 \le 1 - e^2 - \frac{\left( 1 - 6e \right)^2}{2}
  \end{equation}
The above a equation applies when the error rate is in the range
$[0,6/38]$.  The point $e=6/38$ is the point at which the above
equation is maximized. When the error rate exceeds this value the
collision probability saturates. There is no attack which allows
Eve to have complete information on the key. This is in contrast
to BB84 where Eve can steal Alice's photons and send an
uncorrelated photon to Bob. After the measurement basis is
revealed, Eve learns the bit but simultaneously induces a $50\%$
error rate.

Plugging the expression in Eq.~\ref{eq:FinalPcBound} back into
Eq.~\ref{eq:TotalPc}, we obtain the following expression for Eve's
total collision probability on the $k$ bit string,
  \begin{equation}
    Pc = Pc_0^{k\left(1-2\nbar\right)}
  \end{equation}
Using the methods of generalized privacy amplification, the length
of the final key should be set to
  \begin{equation} \label{eq:FinalKeyLength}
    r=-\log_2 Pc - \kappa - s
  \end{equation}
where $\kappa$ is the number of bits exchanged during error
correction and $s$ is a security
parameter~\cite{BennettBrassard95}. The final communication rate,
defined as $R=\lim_{k\to\infty}r/k$, is given by
  \begin{equation}
    R_{DPS} = -p_{click}\left[-\left(1-2\nbar\right)\log_2 Pc_0(e) +
    f(e)h(e)\right]
  \end{equation}
In the above equation $p_{click}$ is the probability Bob detects a
photon, $h(e)=-e\log_2 e - (1-e)\log_2 (1-e)$, and $f(e)$ is a
function which characterizes how far above the Shannon limit the
error correction algorithm is performing
(see~\cite{Lutkenhaus00}).  For error correction algorithms
working in the Shannon limit, which is the ultimate performance
limit of all error correction algorithms, we have $f(e)=1$.

\section{Comparison of DPSQKD to BB84}

Having derived a bound on the average collision probability in the
previous section, we can now compare DPSQKD to the BB84 protocol.
A bound on the collision probability for the BB84 protocol for
realistic sources against individual attacks has been previously
derived in~\cite{Lutkenhaus00}.  In this work, the communication
rate was shown to be
  \begin{eqnarray}
    R_{BB84} & = & p_{click}\left[ -\beta \log_2 \left( \frac{1}{2} + 2\left(\frac{e}{\beta}\right)-2\left(\frac{e}{\beta}\right)^2\right)\right.
     \nonumber \\
    &  & \left. - f(e)h(e)\right]
  \end{eqnarray}
where
  \begin{equation}
    \beta = \frac{p_{click} - p_m}{p_{click}}
  \end{equation}
In the above expression, $p_m$ is the probability that the source
emits a multi-photon state into the channel.

Bob's detection events originate from two sources, the photons
injected into the channel by Alice and dark counts in Bob's
detector.  We assume that both the signal and dark count detection
probabilities are small, so that multiple detection events can be
ignored.  Thus,
  \begin{equation}
    p_{click} = \nbar T + d
  \end{equation}
where $\nbar$ is the average number of photons injected into the
channel, $T$ is the channel transmission, and $d$ is the detector
dark count rate.  The error rate $e$ is given by the expression
  \begin{equation}
    e = \frac{\mu p_{click} +d/2}{p_{click}}
  \end{equation}
where $\mu$ is the baseline error rate of the system due to
imperfections in state preparation, channel induced noise, and
imperfect detection apparatus.

We compare DPSQKD to BB84 using both a Poisson photon source and
ideal single photon source.  For poisson light sources, $\nbar$ is
freely adjustable and $p_m\le \nbar^2/2$.  In contrast, an ideal
single photon source is characterized by $\nbar=1$ and $p_m=0$.
The detector dark count rate is an important parameter in the
simulation.  For telecom wavelengths, one of the most promising
photon detectors is based on up-conversion of $1.5\mu$ photons to
visible wavelengths, where they can be detected using conventional
silicon APDs~\cite{LangrockDiamanti05}.  Such detectors have
already been used to experimentally demonstrate DPSQKD in the
telecom wavelengths, allowing communication distances over 100km
of fiber~\cite{TakesueDiamanti05}.  The experimentally measured
dark count rate for these detectors is 10kHz per detector.  The
APDs have a temporal resolution of 0.5ns.  If the signal is
windowed to this resolution level, the dark count rate per pulse
is $5\times 10^{-6}$ dark counts per detector.  Since DPSQKD uses
2 detectors, the overall dark count rate is $10^{-5}$.  In
contrast, BB84 with passive modulation~\cite{Lutkenhaus99} uses
four detectors giving a dark count rate of $2\times 10^{-5}$. The
baseline error rate is set to $\mu=0.01$.  The parameter $\nbar$
is freely adjustable for BB84 with poisson light, as well as for
DPSQKD.  In the simulations, the value of $\nbar$ is numerically
optimized for each value of the channel loss.

\begin{figure}
\centering\includegraphics[width=7cm]{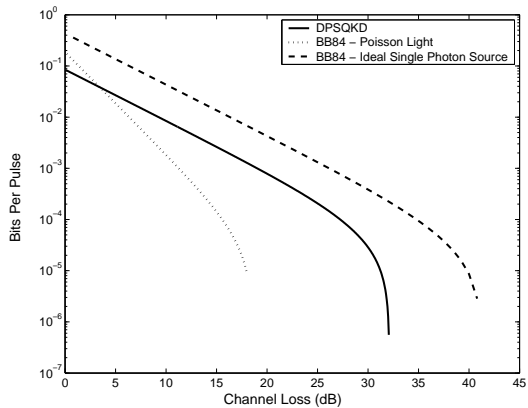}
\caption{Communication rate vs. channel loss for DPSQKD and BB84.}
\label{fig:ratecurve}
\end{figure}

The results of the simulation are shown in
Fig.~\ref{fig:ratecurve}.  The communication rate is plotted vs.
the channel loss in units of dB.  One can see that all three
curves feature an exponential decay for a period of time, after
which the communication rate quickly drops to 0.  This sharp
cutoff is caused by the dark counts in Bob's detectors.  The curve
for BB84 with poisson light decays as a faster exponential than
both DPSQKD and BB84 with an ideal single photon source.  This is
due to photon splitting attacks, which require us to lower $\nbar$
with increasing channel loss.  DPSQKD does not suffer from these
types of attacks, therefore it follows more closely the curve for
BB84 with an ideal single photon source. This is a very important
conclusion, because DPSQKD can be implemented with conventional
lasers, detectors, and linear optics, in contrast to engineering
of ideal single photon sources for BB84.

\section{sequential attacks}

In the previous two sections we investigated the security of
DPSQKD against individual attacks.  The fundamental assumption in
this analysis was that Eve measures each photon independently, and
does not use the measurement results of some of the photons to
refine the measurement of the remaining photons.  However, in
DPSQKD there are certain attacks which do not satisfy this
assumption, but which are conceptually very simple. One such
attack is the sequential attack.

In a sequential attack, Eve uses a detection apparatus equivalent
to Bob's setup, which she places in the quantum channel very close
to Alice.  Eve then waits for $k$ consecutive clicks on her
detection apparatus.  Whenever such an event occurs, Eve can
reconstruct a $k+1$ time slot state.  This states induces an error
rate of
  \begin{equation}
    \epsilon_{seq} = \frac{1}{2(k+1)}
  \end{equation}
Off course, the probability of observing $k$ consecutive clicks
decreases exponentially with $k$.  If $\nbar$ is the average
number of photons per pulse, then the probability of $k$
consecutive clicks is $\nbar^k$.  This probability must be at
least as large as Bob's detection probability in order for Eve to
conserve the overall detection rate.  Thus, we must have
$\nbar^k\ge\nbar T$, which imposes an upper bound on $k$.

The collision probability for sequential attacks is very easy to
calculate.  When Bob detects a photon in any time slot other than
slot $1$ or $k+2$, Eve knows the value of Alice's key.  This
happens with probability $k/(k+1)$.  If Bob detects a photon in
slot $1$ or $k+2$, then Eve knows nothing about Alice's key, so
her collision ptobability is 1/2.  If Eve performs $M$ sequential
attacks, her collision probability is given by
  \begin{equation} \label{eq:PCSeq}
    P_{c0} = \frac{1}{2^{M/k+1}}
  \end{equation}
From the condition $\nbar^k=\nbar T$ we obtain that
  \begin{equation}
    k = \log_{\nbar}T + 1
  \end{equation}
This condition ensures that there are enough sequential clicks to
conserve the communication rate.  However, even if the number of
sequential clicks is sufficient, Eve may not be able to perform an
attack on every bit of the key, because she cannot exceed the
natural system error rate which we define as $\epsilon_s$.  She
can only perform a sequential attack on a fraction $\epsilon_s /
\epsilon_{seq}$ of the bits, and must leave the remainder of the
string undisturbed to conserve the error rate.  Thus, if $N$ is
the number of bits in Alice's string, then
  \begin{equation}
    M = \frac{N\epsilon_s}{\epsilon_{seq}} =
    N(k+1)\epsilon_s
  \end{equation}
Plugging the above equation into Eq.~\ref{eq:PCSeq}, and using
Eq.~\ref{eq:FinalKeyLength}, we obtain the communication rate
  \begin{equation}
    R_{seq} = p_{click} \left[ 1 - 2\epsilon_s \left(
    \log_{\nbar}T +1 \right) - f(e) h(e) \right]
  \end{equation}

\begin{figure}
\centering\includegraphics[width=7cm]{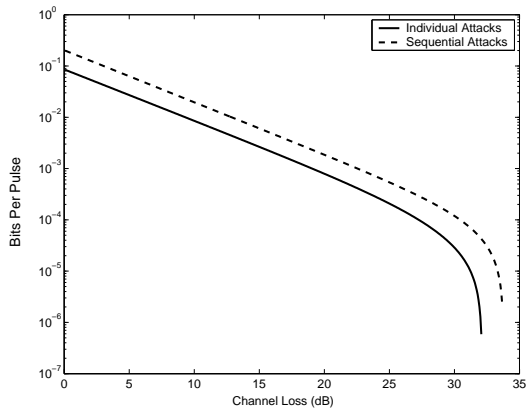}
\caption{Comparison of individual attacks to sequential attacks in
DPSQKD.} \label{fig:SeqRate}
\end{figure}

We compare this communication rate to that of DPSQKD calculated in
the previous section.  Using the same values for the dark count
and error rate, we plot the communication rate for sequential
attacks and individual attacks in Fig.~\ref{fig:SeqRate}.  For
individual attacks, the average photon number $\nbar$ is once
again optimized for each value of the channel loss.  We then use
the same optimal $\nbar$ to evaluate the rate for sequential
attacks, so that we may compare the effectiveness of individual
and sequential attacks under the same operating condition.  One
can see that the communication rate for individual attacks is
always lower than sequential attacks, indicating that in the
operating regime we are considering it is more advantageous for
Eve to perform individual instead of sequential attacks.  This
means that security against individual attacks already implies
security against sequential attacks as well.

Off course, we do not know if the sequential attack are optimal,
or if a more clever scheme could produce better results for Eve.
To answer this question, a more general proof of security is
needed.

\section{conclusion}

In conclusion, we have derived a proof of security for DPSQKD with
realistic sources against individual attacks.  This proof allows
us to directly calculate the communication rate after privacy
amplification.  We showed that, in contrast to BB84, DPSQKD does
not suffer from photon splitting attacks even when implemented
with attenuated lasers.  We compared the communication rate as a
function of channel loss for DPSQKD to BB84 using both an
attenuated laser and ideal single photon source.  DPSQKD allows us
to achieve communication rates close to BB84 with an ideal single
photon source, making it an outstanding candidate for practical
long distance quantum cryptography.  We then compared individual
attacks to sequential attacks in DPSQKD and showed that individual
attacks are more powerful in our operating regime.  Thus, security
against individual attacks already ensures security against
sequential attacks as well.

Financial support for this work was provided by the MURI Center
for photonic quantum information systems (ARO/ARDA Program
DAAD19-03-1-0199), as well as a DCI fellowship.

\appendix

\section{Expression for collision probability}\label{ap:Pc}

Here we derive the expression for the collision probability given
in Eq.~\ref{eq:IndPcRearranged}.  We start with
Eq.~\ref{eq:IndPc}, and use Bayes rule to rewrite it as
  \begin{equation}
    Pc_0 = \sum_m p(m) \sum_z
    \frac{p^2(z|0,m)p^2(0|m) + p^2(z|1,m)p^2(1|m)}
    {p(z|m)}
  \end{equation}
By completing the square, we can re-write the above expression as
  \begin{equation}
    Pc_0 = \sum_m p(m) \left( 1 - 2\sum_z
    \frac{p(0) p(1) p(z,m|0)p(z,m|1)}
    {p(z,m) p(m))} \right)
  \end{equation}
Using the fact that $p(0)=p(\pi)=1/2$ directly leads to the result
stated in Eq.~\ref{eq:IndPcRearranged}.

\section{Derivation of the error rate}\label{ap:ErrRate}

In this section we show that Eve's attack strategy leads to an
error rate given by Eq.~\ref{eq:ErrRateTot}.  We start with the
obvious relation $p_{e,m} = (p_{e,m|0}+ p_{e,m|1})/2$.  We define
the states $\|M_+>=\|J_m> + \|J_{m+1}>$ and $\|M_->=\|J_m> -
\|J_{m+1}>$.  We define $E_{\phi_1\ldots\phi_k}[A]$ as the average
of expression $A$ over the possible values of
$\phi_1\ldots\phi_k$.  It is straightforward to show that
  \begin{eqnarray*}
    p(m) & = & \frac{1}{4N}E_{\phi_1\ldots\phi_k}\left[ \bk<M_-|M_-> +
    \bk<M_+|M_+> \right] \\
    & = & \frac{1}{2N} \sum_n \bk<E_{n,m}|E_{n,m}> +
    \bk<E_{n,m+1}|E_{n,m+1}>
  \end{eqnarray*}

Now,
  \begin{eqnarray*}
    \lefteqn{p_{e,m|0}   =
    \sum_{\phi_1,\ldots,\phi_k}p_{e,m|0,\phi_1,\ldots,\phi_k}\prod_{j\ne m+1}p(\phi_j)} \\
    & = &
    \sum_{\phi_1,\ldots,\phi_k}p_{e,m|0,\phi_1,\ldots,\phi_k}2^{-(k-1)}
    \\
    & = & \sum_{\phi_1,\ldots,\phi_k} \bk<M_-|M_-> 2^{-(k-1)}\\
    & = & \frac{1}{4N} \sum_{n \ne m, m+1}
    |\|E_{n,m}>-\|E_{n,m+1}>|^2 + \\
    & & |\left(\|E_{m,m}>-\|E_{m+1,m+1}>\right) + \left(\|E_{m+1,m}>-\|E_{m,m+1}>\right) |^2
  \end{eqnarray*}
The exact same argument leads to
  \begin{eqnarray*}
     \lefteqn {p_{e,m|0}  =  \frac{1}{4N} \sum_{n \ne m, m+1}
    |\|E_{n,m}>-\|E_{n,m+1}>|^2 +} \\
      & &|\left(\|E_{m,m}>-\|E_{m+1,m+1}>\right) - \left(\|E_{m+1,m}>-\|E_{m,m+1}>\right)
      |^2
  \end{eqnarray*}
Using the above two expressions we have
  \begin{eqnarray*}
    p_{e,m} & = & \frac{1}{2} \left[ p(m) -  \frac{1}{N}\left(
    \bk<E_{m,m}|E_{m+1,m+1}> \right.\right. \\
    & & \left.\left.  + \bk<E_{m+1,m}|E_{m,m+1}> \right)\right]
  \end{eqnarray*}
Dividing the above expression by $p(m)$ directly leads to the
expression in Eq.~\ref{eq:ErrRateTot}.

\section{Expression for collision probability}\label{ap:CollProb}

Here we derive the expression in Eq.~\ref{eq:PcRaw}.  We start
with the expression in Eq.~\ref{eq:IndPcRearranged}.  Using the
same definition for $E_{\phi_1\ldots\phi_k}[A]$ that we did in
appendix~\ref{ap:ErrRate}, we have
  \begin{eqnarray*}
    p(z,m|0) & = &\frac{1}{4N}E_{\phi_1\ldots\phi_k}\left[ |
    \left(\bk<z|J_m>+\bk<z|J_{m+1}>\right)\|0_m>  \nonumber \right.\\
    & & \left.+\left(\bk<z|J_m>-\bk<z|J_{m+1}>\right)\|1_m>|^2 \right] \\
    & = &\frac{1}{4N}\left[
    \left( E_{m,m}(z) + E_{m+1,m} \right)^2 \right. \\
    & &+\left( E_{m,m+1}(z) + E_{m+1,m+1}^2 \right)^2 \\
    & &\left.+ \sum_{n\ne m,m+1} E_{n,m}^2 + E_{n,m+1}^2 \right]
  \end{eqnarray*}
Similarly we can derive
  \begin{eqnarray*}
    p(z,m|1) & = &\frac{1}{4N}\left[
    \left( E_{m,m}(z) - E_{m+1,m} \right)^2 \right. \\
    & &+\left( E_{m,m+1}(z) - E_{m+1,m+1} \right)^2 \\
    & &\left.+ \sum_{n\ne m,m+1} E_{n,m}^2 + E_{n,m+1}^2 \right]
  \end{eqnarray*}
Using the fact that $p(z,m)=(p(z,m|0) + p(z,m|1))/2$, and plugging
the above two expressions into Eq.~\ref{eq:IndPcRearranged}
directly leads to the expression given in Eq.~\ref{eq:PcRaw}.

\section{Upper bound on collision probability}
\label{ap:ColProbBound}

We start with equation~\ref{eq:PcRaw}, and use the form of the
Cauchy inequality which was first proposed by Lutkenhaus for the
bound on the collision probability in BB84 (see Appendix A
of~\cite{Lutkenhaus99} ).  Specifically if $\psi(z)=\bk<z|\psi>$
and $\phi(z)=\bk<z|\phi>$, then the Cauchy inequality tells us
that
\begin{widetext}
  \begin{equation}
    \sum_z
    \frac{\psi^2(z)\phi^2(z)}{A_m^2(z)+A_{m+1}^2(z)+B_m^2(z)+B_{m+1}^2(z)+C_m^2(z)+C_{m+1}^2(z)}
    \ge \frac{\bk<\phi|\psi>}{2p(m)}
  \end{equation}
\end{widetext}
We expand the product terms in Eq.~\ref{eq:PcRaw}, and apply the
above bound.  Also, we can assume that $\|A_m>$ and $\|C_{m+1}>$
are orthogonal to all other vectors, because this maximizes the
collision probability without affecting the error rate.  This
leads directly to the expression given in Eq.~\ref{eq:PcBound0}.

\section{Symmetrization of collision probability}
\label{ap:SymAttack}

We have so far shown that the collision probability and error rate
depend on interference between state vectors at times $m$ and
$m+1$.  This means that our optimization problem has a symmetry of
circular permutation.  Specifically, if we apply the following
transformation,
  \begin{eqnarray*}
    \|A_m> & \to & \|A_{m+1\mod k}> \\
    \|B_m> & \to & \|B_{m+1\mod k}> \\
    \|C_m> & \to & \|C_{m+1\mod k}>
  \end{eqnarray*}
we do not affect the error rate or Eve's collision probability.
Now, let us suppose that an optimal attack exists which is given
by the state vectors $\|A_m>$, $\|B_m>$, and $\|C_m>$.  We can
form a new set of state vectors $\|A'_m>$, $\|B'_m>$, and
$\|C'_m>$ as follows
  \begin{eqnarray*}
    \|A'_m> & = & \frac{1}{\sqrt{k}} \sum_{j=0}^{k-1} \|A_{m+j\mod
    k}>\|j> \\
    \|B'_m> & = & \frac{1}{\sqrt{k}} \sum_{j=0}^{k-1} \|B_{m+j\mod
    k}>\|j> \\
    \|C'_m> & = & \frac{1}{\sqrt{k}} \sum_{j=0}^{k-1} \|C_{m+j\mod
    k}>\|j> \\
  \end{eqnarray*}
In the above equations, $\|j>$ represent an orthogonal basis which
keeps track of which circular permutation has been chosen.  The
collision probability can now be written as
  \begin{eqnarray*}
    P_{c0} & = & \sum_{x,z,m,j} p^2(x|z,m,j)p(z,m,j) \\
    & = & \sum_j p(j) \sum_{x,z,m} p^2(x|z,m,j)p(z,m|j) \\
    & = & \sum_j p(j) P_{c0|j}
  \end{eqnarray*}
The expression $P_{c0|j}$ is simply the average collision
probability given the value of the measurement on the states
$\|j>$.  However, because the different values of $j$ represent
different circular permutations and the collision probability is
invariant under circular permutation, we have $P_{c0|j}=P_{c0}$.
Thus, the symmetrized probes $\|A'_m>$, $\|B'_m>$, and $\|C'_m>$
have the same collision probability as the un-symmetrized ones. It
is easy to verify that these symmetrized probes satisfy the
property that their inner products with each other is independent
of $m$.

\section{Optimization of the collision
probability}\label{ap:optimization}

We define $a=\bk<A_0|A_0>=\bk<A_1|A_1>$,
$b=\bk<B_0|B_0>=\bk<B_1|B_1>$, and $c=\bk<C_0|C_0>=\bk<C_1|C_1>$.
Normalization imposes the constraint $a+b+c=1$.  We define the
angles $\phi_1$ and $\phi_2$ as
  \begin{eqnarray*}
    \bk<B_1|B_0> & = & b \cos \phi_1 \\
    \bk<A_1|C_0> & = & \sqrt{ac} \cos \phi2
  \end{eqnarray*}
Straightforward manipulation of the bound on $P_{c0}$ leads to the
expression
  \begin{eqnarray*}
    \lefteqn{P_{c0} \le 1 - \frac{1}{8}  \left[ a^2 + c^2 + (b-c)^2 + \right.}\\
    & &  \left. (b-a)^2 + 2 (b\cos \phi_1 - \sqrt(ac)\cos\phi_2)
    \right]
  \end{eqnarray*}
We also use the fact that
  \begin{equation*}
    \left( b \cos \phi_1 - \sqrt{ac}\cos\phi_2 \right) =
    \left(1-2e\right)^2 - 4 b\sqrt{ac}\cos\phi_1\cos\phi_2
  \end{equation*}
Using the above expression, it is easy to show that the collision
probability is maximized and the error rate is minimized when
$\cos\phi_1 = \cos \phi_2 = 1$.

Now we set
  \begin{eqnarray*}
    a & = & (1-b) \cos \theta \\
    c & = & (1-b) \sin \theta
  \end{eqnarray*}
Plugging into the expression for the collision probability, it is
straightforward to show that the collision probability achieves a
maximum when $\theta = \pi/4$, and that this condition also
minimizes the error rate.  Thus, the optimal attack strategy
occurs when $a=c$.  This condition implies that
  \begin{eqnarray*}
     e & = & \frac{x}{2} \\
     P_{c0} & \le & 1 - \frac{1}{4} \left(x^2 + 2(1-3x)^2 \right)
  \end{eqnarray*}
Substituting the expression for $e$ into $P_{c0}$ directly leads
to the expression in Eq.~\ref{eq:FinalPcBound}.

\end{document}